\documentstyle[12pt]{article}

\title{\large ROLE OF INTERFACES IN THE PROXIMITY EFFECT
IN ANISOTROPIC SUPERCONDUCTORS }

\author{E. Polturak, G. Koren, D. Cohen, O. Nesher, \\
Physics Department\\
Technion - Israel Institute of Technology, Haifa 32000, ISRAEL,\\
and\\
R. G. Mints and I. Snapiro\\School of Physics and Astronomy,\\
Tel Aviv University, Tel Aviv 69978, ISRAEL}
\date{}
\begin{document}
\normalsize
\baselineskip=7mm
\maketitle
\centerline{ABSTRACT}
We report measurements of the critical temperature of 
YBCO-$\rm YBCu_{3-x}Co_xO_{7+\delta}$Superconductor-Normal bilayer films. 
Depending on the morphology of the S-N interface, the coupling between 
S and N layers can be turned on to depress the $\rm T_c$
of S by tens of degrees, or turned down so the layers appear 
almost totally decoupled. This novel effect can be explained 
by the mechanism of quasiparticle transmission into an 
anisotropic superconductor.

\vspace{15mm}
\pagebreak
The system of a high $\rm T_c$ superconductor and a normal conductor 
in proximity received much attention, not least since one of the practical 
high $\rm T_c$ Josephson junction types is the 
Superconductor-Normal-Superconductor (SNS) device[1]. 
Despite the effort spent by many groups,these junctions almost 
invariably behave as if the normal barrier contains many 
superconducting microshorts, even when the thickness of the normal layer 
is several hundred $\rm \AA$,and should be totally pinhole free. It 
occured to us that this effect may not be connected with the properties 
of the S or N layers per se, but rather due to some fundamental physics 
of the transport across the S/N interface. To check this idea experimentally, 
we decided to compare the superconducting properties of S/N bilayers 
differing only in their interfaces. We know how to control the morphology 
well in a configurationof a {\bf c} axis oriented films. The bilayer films 
consist of a thin layer of YBCO grown on (100) $\rm SrTiO_3$, capped by a 
much thicker layer of $\rm YBCu_{3-x}Co_xO_{7+\delta}$. 
$\rm YBCu_{3-x}Co_xO_{7+\delta}$ was chosen as the normal material since 
it grows epitaxially on YBCO, has a negligible interdiffusion, and was used 
as a barrier in SNS junctions by several groups [2,3].
The expected influence of the normal conductor on the superconductor 
would extend to a distance of a coherence length from the interface, as 
predicted by the conventional proximity effect[1]. Accordingly, the S layer 
was much  thinner than the N layer in order to produce an observable effect 
on$\rm T_c$ of the bilayer.\\ 

The films, deposited using laser ablation, are epitaxial with {\bf c} axis 
perpendicular to the substrate. Bilayers were prepared with the thickness 
of YBCO between 60$\rm \AA$ and 550$\rm \AA$, and a thicker 
$\rm YBCu_{3-x}Co_xO_{7+\delta}$ cap, between 1000$\rm \AA$and 1500$\rm \AA$. 
Here we present  data for
$\rm YBCu_{3-x}Co_xO_{7+\delta}$ with x=0.15, which is normal down to 
about 50K.To maintain a clean interface between the S and N layers, 
each bilayer was grown in one deposition run. Different S/N interfaces 
were produced by changing the deposition rate. We refer to a rate of 
deposition of $\rm 6.5\AA /sec$ as fast growth, while bilayers grown at a 
rate 2.5 times slower,  with the growth interrupted every 30 sec. of 
deposition for a 60 sec. pause, we call slowly grown. The fast growth 
rate produces films which grow by screw dislocations and their surface 
shows rounded smooth features, similar to those obtained by 
Schlom et al.,[4]. In contrast, the slow growth produces films which grow 
in the Stranski-Krastanov mode, namely  layer by layer, up to a critical 
thickness of about $\rm 150\AA$, and then by a 2D island growth[5]. 
The bilayers were characterized by transport, ac susceptibility and 
Atomic Force Microscopy (AFM). Overall, the reproducibility of data 
was excellent over a period of a year or so during which the experiments 
were done. A summary of the $\rm T_c's$ of all these bilayers plotted 
against the thickness of the YBCO layer, is shown in Fig. 1. In order 
to have only one variable, we chose to work in the regime where
$\rm T_c$ of the bilayers becomes independent of the thickness of the 
cap layer, and thus independent of the top 
surface of bilayer. One can see in the inset of Fig. 2, that this happens 
once the thickness of the cap exceeds about 1000$\rm\AA$. All the data 
presented
here were obtained in this regime. One can see that there is an enormous 
difference between the $\rm T_c's$ of the fast grown ( solid symbols) and 
the slowly grown ( open symbols ) bilayers. In contrast, the difference 
between the $\rm T_c's$ of single layer YBCO films grown as reference at
these two deposition rates were no more than 2-3K, as shown in Fig. 2.  
Furthermore, $\rm T_c$ of single layer fast grown films as thin as 
$\rm 100\AA $ is near 85K, which shows that there are no problems with 
the films of this thickness being discontinous. 
Thus, the lowering of $\rm T_c$ of the bilayers is definitely associated 
with the presence of the normal layer. \\    

Since the effect was much larger than expected, our first thought was that 
the reduction of $\rm T_c$ is caused by migration of oxygen from the YBCO 
into $\rm YBCu_{3-x}Co_xO_{7+\delta}$, leaving the YBCO oxygen deficient, 
and hence with a lower $\rm T_c$. Fully oxygenated 
$\rm YBCu_{3-x}Co_xO_{7+\delta}$ contains more oxygen than 
YBCO, with the excess amount increasing with x[6]. If not enough oxygen 
is supplied during growth,then some oxygen may subsequently migrate 
from the YBCO into the $\rm YBCu_{3-x}Co_xO_{7+\delta}$, lowering the 
$\rm T_c$ of the YBCO. To check this possibility, we grew bilayers 
under different methods of oxygen loading as follows: (a): increased 
the oxygen ambient pressure during deposition by a factor of 2. 
(b): increased the time length ofthe post deposition oxygen loading by 
a factor of 2. (c): increased the Co composition of the capping layer 
from x=0.15 to x=0.3 and then to x=1. Test (a) increases the flux of oxygen 
atoms during growth. Test (b) allows the film to absorb more oxygen during 
post deposition oxygen loading. Both (a) and (b) should increase the amount 
of oxygen in the film, and thus increase $\rm T_c$. In contrast, test (c) 
increases the amount of excess oxygen needed by 
$\rm YBCu_{3-x}Co_xO_{7+\delta}$ 
over that of YBCO[6],so if the oxygen migrates from the YBCO into 
$\rm YBCu_{3-x}Co_xO_{7+\delta}$, $\rm T_c$ should be further lowered. The 
results were that $\rm T_c$'s of the bilayers made in these various methods 
were the same. Thus, oxygen doping is not the reason of the reduction 
of $\rm T_c$. \\  

The mutual influence of N and S as expressed in Figs. 1 and 2 is 
reminiscent
of the conventional proximity effect[7]. 
We first consider which of our observations are generally consistent 
with this
picture and which are not. To begin, we discuss the ''saturation thickness'',
namely the thickness of N above which the influence of N on S saturates 
(see inset of Fig. 2). In the proximity effect, $\rm T_c$ 
of a bilayer reflects the
balance between the number of quasiparticles transmitted from N into S
and pairs transmitted in the opposite direction. 
Pairs penetrate into N a distance of several normal coherence lengths,
$\xi_N$. It is therefore plausible 
that the depth from which quasiparticles in N will reach the
interface and penetrate into S should be quite similar. 
Values of $\xi_N$ of 270$\rm \AA$ have been measured for example 
in {\bf c}-axis oriented PrBCO films[8], so that the $\rm \sim 1000\AA$ 
that we 
find as the thickness of N at which the influence of N on S saturates appears 
to be within several such $\xi_N$. Thus, the ''saturation thickness'' of N
seems within the bounds of the conventional picture.
However, three other observations
reported here do not fit the conventional proximity effect. First, the 
depression of $\rm T_c$ of the bilayers does not depend on the Co doping 
level of the $\rm YBCu_{3-x}Co_xO_{7+\delta}$, and hence on the $\rm T_c$ of
the normal material. This effect is not understood at present. Second, the 
depression of $\rm T_c$ of the bilayers seems to be very different for
fast grown and slowly grown bilayers. Third, a large
depression of $\rm T_c$'s of 
the fast grown bilayers occurs for YBCO layers much thicker than the 
coherence length of S. The remainder of the paper is devoted to 
discussing possible explanations of the last two effects mentioned.

We first turn to discuss the reason for the difference between the slowly
grown and fast grown bilayers. The morphology of the surface
of slowly and fast grown single layer films, measured by AFM, is shown 
in Fig. 3( these surfaces are the interface in the bilayers ).There is 
indeed a striking difference between the features visible on the surface of 
the fast grown film, which are rounded and isotropic, while the slowly grown 
film shows a very regular array of pyramids, reflecting the symmetry of 
the underlying lattice. Despite the difference in the details, the averaged 
interface roughness of the fast grown and slow grown bilayers is quite 
similar. For example, the rms roughness of a slowly grown film 250$\rm \AA$ 
thick is 24$\rm \AA$, while for a fast grown film of the same thickness it 
is 30$\rm \AA$. Consequently, if all that mattered for the coupling 
between S and N layers was the area of the interface, the reduction of 
$\rm T_c$ should be very close for the two types of films. Clearly, this is
not the case. We propose that the difference between these two types
is unique to anisotropic superconductivity.\\ 

In the context where S and N are in proximity, the decrease of $\rm T_c$ 
of the S layer is due to the presence of excess number of normal 
quasiparticles, transmitted from N into S [7]. When the order parameter is 
isotropic, the orientation of the interface is not an important factor in 
the transmission of quasiparticles from N into S. However, in the case of 
anisotropic order parameter, (d-wave, or s+d) Tanaka and Kashiwaya[9] 
have predicted that the transmission coefficient is strongly dependent on 
$\theta$, the angle between the high symmetry 
crystalline directions ( to which the order parameter is locked ) and the 
normal to the interface. Essentially the same conclusion was 
reached by Barash, et al. [10].  In particular, the transmission through 
the interface should be  anomalously large in the directions along which 
the order parameter has a minimum ( $\theta$=$\pi/4$, where interface normal 
is parallel to the diagonal between {\bf a} and
{\bf b} in the case of d-wave, or s+d order parameter ).
Depending on the value of interfacial potential barrier, the transmission 
along these diagonal directions can be tens of times larger than along 
the high symmetry directions [11]. It was shown recently that this effect is
responsible also for the Zero Bias Anomalies found in high $\rm T_c$ tunnel
junctions[12]. On the other hand,
when the interface normal is parallel to {\bf a} or {\bf b}, 
the transmission will be similar to that in isotropic superconductivity. 
The anisotropy of the transmission survives the summation
over all angles at which quasiparticles are incident on the
interface, because their pair potential is symmetric
about the crystalline axes, and 
therefore is not symmetric about the normal to the interface, which
means the result of the summation will depend on $\theta$, 
unless this normal and one of the crystalline high 
symmetry directions happen to coincide[11].  On the basis of these models, 
two additional statements can be made regarding the
the interfaces in our bilayers. First,  
Barash, et al.[10], argue that the anisotropy would survive the
averaging over all incidence angles
only when the interface scattering is specular. 
On the basis of the differences between the slowly grown bilayers and
the fast grown ones, this condition is fulfilled
by the interfaces in our experiment. Second, 
the transmission through the interface[9,11] depends linearly on 
$\theta$, being low for small values of $\theta$
and maximum for $\theta$=$\pi/4$. The large difference between the
fast grown and slowly grown bilayers arises since most of the of the 
interface of the slowly grown bilayers is oriented with $\theta$ close to 
zero, as seen in Fig. 3, while in the fast grown bilayers all crystalline 
directions are equally exposed. Therefore, the transmission through the 
interface of the fast grown bilayers is much larger and their $\rm T_c$ 
would correspondingly be lower. We find it remarkable that just by
changing the morphology of the surface the film can switch from
normal to anomalously high transparency. Our data therefore illustrate 
rather vividly the crucial role of anisotropic superconductivity. \\

We now discuss the question of the length scale inside S where $\rm T_c$ 
is reduced.The reduction of $\rm T_c$ for a bilayer, $\Delta T_c$,  
expected due to the proximity effect is given by 
$\Delta T_c/T_{c0}\approx -1.35\,\xi_c^2(0)/d_S^2$, where $d_S$ is the 
thickness of the superconducting layer
and $\rm T_{c0}$ is the transition temperature of a single layer thick 
film[7]. In the case of a perfectly smooth, {\bf c} axis oriented film, 
$\xi_c(0)\sim\rm 3\,\AA $, and the predicted decrease of $T_c$ of a 
$\rm 100\,\AA$ film is about 0.1K. If the S/N interface is not planar, 
some in-plane coupling will be present, extending the range of the influence 
of N on S. To obtain a quantitative comparison, we extended the calculation 
of $\Delta T_c$ for the case when the S/N interface is rough. We use the 
Ginzburg-Landau approach and describe this interface by a function 
$z_S(x,y)=d_S+f(x,y)$, where the directions of x and y are chosen so that 
$\hat{\bf x}\,\|\,{\bf a}$, and  $\hat{\bf y}\,\|\,{\bf b}$. We also assume 
that $|f(x,y)|\ll d_S$, and that the value of $f(x,y)$ averaged over the 
interface surface $\langle f\rangle =0$. In this case $\Delta T_c$ is given 
by
\begin{equation}
{\Delta T_c\over T_{c0}}\approx 1.35\,{\xi_c^2(0)\over d_S^2} 
+0.72\,{\xi_a^2(0)\langle f_a^2\rangle +\xi_b^2(0)\langle f_b^2\rangle 
\over d_S^2},
\end{equation}
where $f_i=\partial f/\partial x_i$. A direct measurement of the
roughness of the film is not viable on the scale of $\xi_a(0)$ 
(or $\xi_b(0)$). However, one can set a limit using the fact that the 
surface of the film is composed of unit cell size steps in the ${\bf c}$ 
direction on planar terraces. The density of steps on the surface can be 
calculated from the macroscopic inclination of the film surface relative to 
the substrate. Using the AFM pictures of the films such as shown in Fig.~2, 
we calculated $\langle f_a^2\rangle$ as a function of $d_S$. For example, 
for $d_S={\rm 500\AA}$ we find $\langle f_a^2 \rangle\rm{\sim 1.3}$ for a 
fast grown film and 0.5 for the slow grown one. For 
$\langle f_a^2\rangle\sim\rm 1$, the second term in Eq.~(1) is much bigger 
than the first one. The result of the calculation of 
$\rm T_c = T_{c0} -\Delta T_c$ is shown in 
Fig.~1 as the solid line. We took for this calculation 
$\xi_a(0)=\xi_b(0)=\rm 20\,\AA$ and $\rm T_{c0}= 90\,K$. We find that if 
one takes the interface roughness into account, then
the conventional proximity effect is consistent, without any adjustable
parameters, with the dependence of $T_c$ on $d_S$ for 
the slowly grown bilayers.
This result is also consistent with the transmission 
through the interface being normal in this case, 
namely similar to that found in isotropic 
superconductors. However, the proximity effect cannot account  
for the data of the fast grown bilayers, where the transmission
through the interface is strongly enhanced. It is therefore an open
question whether the fast grown and slow grown films should be treated
on the same footing ( in this case, the proximity effect does not work)
or should the fast grown films be described by a totally different
theory.\\

In equilibrium, the length scale describing  the influence of N on S 
is $\sim\xi_0$. Looking at $\rm T_c$'s of the
fast grown bilayers in Fig. 1, one can see a crossover between the
thin film regime, where $\rm T_c$ is very close to that of  
$\rm YBCu_{3-x}Co_xO_{7+\delta}$, about 50K for x=0.15, and
the thicker film regime, where $\rm T_c$ approaches that of YBCO. 
This crossover takes place at $d_S\sim {\rm 300 \AA}$, which clearly
is much larger than $\xi_0$. This is essentially why the proximity effect
fails to describe this case. We know of no theory pertinent to these
experimental observations. However, we may mention that in the context of 
the usual s-wave superconductivity, 
Blonder et al.[13] considered another characteristic length, that for the
direct conversion of a current of normal quasiparticles entering S into 
pairs. This  diffusion length is given by $\Lambda_Q=\sqrt{D \tau_Q}$. 
Here, $D$ is the diffusion coefficient and $\tau_Q$ is the quasiparticle 
lifetime with respect to recombination. We calculated this length both 
for in-plane and out of plane diffusion. In the plane, $D_{ab}=v_Fl/3$, 
where $v_F$ is the Fermi velocity, and $l$ is the mean free path. 
Taking $v_F={\rm 5\times 10^7}$ cm/sec[14], $l\sim{\rm6 \times 10^{-7}}$cm, 
and $\tau_Q$=5 psec [15], we find the in plane value of $\Lambda_Q(a-b)$ 
is about 700$\rm\AA$.  Regarding the {\bf c} direction, quasiparticles 
diffuse mainly via interlayer scattering with a diffusion coefficient 
$D_c\sim c^2/t_c$[16]. Here, $c={\rm 12\AA}$ is the interlayer distance and 
$1/t_c$ is the scattering rate. Taking $t_c={\rm 1.5\times 10^{-14}}$ sec[16], 
we find $\Lambda_Q(c){\rm =220\AA}$. Experimentally, $\tau_Q$ is constant[15]
throughout the temperature range of this work, and therefore the diffusion
lengths are also temperature independent. It is evident that the diffusion 
lengths are comparable to the 
experimental crossover length of 300$\rm \AA$.
However, the diffusion length describes
the conversion of normal current into a supercurrent in a BCS s-wave
superconductor[13], whereas our
case brings out out the dramatic difference between the isotropic
and anisotropic superconductivity. Thus, altough the numbers are suggestive, 
it would clearly be imprudent to identify the recombination length 
as describing our data without the having a theoretical calculation
generalizing this particular result of Blonder, et al., to the anisotropic
case. We hope that this work may in fact stimulate interest to do this
calculation.\\

In conclusion, we demonstrated a dramatic effect of the morphology
of the S/N interface on the properties of the bilayers. It is 
painfully obvious that 
one has to consider this effect in the future design of high $\rm T_c$ 
Josephson junctions.\\

We are grateful to L. P. Pitaevskii, D. Artenberg, B. Fisher and 
C. Villard for helpful discussions. This work was supported in part by 
Israeli Science Foundation, The Minerva Foundation, and the Technion Fund 
For promotion of Research. RGM and IS acknowledge the support of the 
German-Israeli Foundationfor Research and Development. 
\newpage

REFERENCES\\
\begin{enumerate}
\item K. A. Delin and A. W. Kleinsasser, Supercond. Sci. and Technology 
{\bf 9}, 227( 1996).
\item B. Moeckley and K. Char, Physica {\bf C265}, 283 (1996), 
L. Antognazza, S. J. Berkowitz, T. H. Geballe, and K. Char, 
Phys. Rev. {\bf B51}, 8560 (1995).
\item G. Koren and E. Polturak, Physica {\bf C230}, 340 (1994).
\item D. G. Schlom et al., Z. Phys. Cond. Mat. {\bf B86}163 ( 1992 ).
\item X. Y. Zheng et al, Phys. Rev. {\bf B45}, 7584 (1992).
\item W. M. Chen, et al., Physica {\bf C 270}, 349 (1996).
\item G. Deutscher and P. G. de-Gennes, in {\em Superconductivity}, Vol. 2, 
R. Parks ed.,( Marcel Dekker, New York, 1969 ).
\item A. Schattke, et al., in {\it Proc. Appl. Sup. 1997}, IOP 
Conference Series \#158 (1997).
\item Y. Tanaka and Kashiwaya, Phys. Rev. Lett. {\bf 74}, 3451 (1995).
\item Yu. S. Barash, A. V. Galaktionov, and A. D. Zaikin, 
Phys. Rev. {\bf 52}, 665 ( 1995).
\item Y. Tanaka, in {\em Coherence in HTS Superconductors}, G. Deutscher and 
A. Revcolevschi editors, ( World Scientific, Singapore, 1996)
\item M. Covington, et al., Phys. Rev. Lett. {\bf 79}, 277 ( 1997), 
and references therein.
\item G. E. Blonder, M. Tinkham, and T. M. Klapwijk, Phys. Rev. {\bf B25}, 
4515 ( 1982 ).
\item M. Weger, Jour. Low Temp. Phys. {\bf 95}, 131 ( 1994 ).
\item C. J. Stevens, et al., Phys. Rev. Lett.{\bf 78}, 2212 (1997).
\item S. L. Cooper and K. E. Gray, in {\em Physical Properties of
High Temperature Superconductors}, D. M. Ginsberg ed., vol. 5, p. 61,
( World Scientific, Singapore, 1994).

\end{enumerate}
\newpage
FIGURE CAPTIONS\\
\\
{\bf Fig. 1.}\ $\rm T_c$ of the S/N bilayers vs. the thickness of the  YBCO.
Closed symbols refer to fast grown bilayers, and open symbols refer to 
slowly grown bilayers. The solid line is $\rm T_c$ expected  from the 
proximity effect in the case of a rough interface.\\

{\bf Fig. 2.}\ $\rm T_c$ of single layer reference YBCO films vs. 
their thickness. Closed squares refer to fast grown films, and open circles 
to slowly grown films. The inset shows the dependence of the $\rm T_c$ of
a bilayer on the thickness of the $\rm YBCu_{2.85}Co_{0.15}O_{7+\delta}$ cap.
The thickness of the (fast grown ) YBCO film in these bilayers is 
250$\rm \AA$. Solid line is a guide to the eye.\\

{\bf Fig. 3.} \ Atomic Force Microscope pictures of the surface of the  
single layer slowly grown film ( a) and the fast grown film (b). 
These surfaces are the interfaces in the bilayer geometry. The 
average thickness of both films is about 500$\rm \AA$, and the scale 
of the features shown represents $\rm \pm 10\%$ thickness variation.

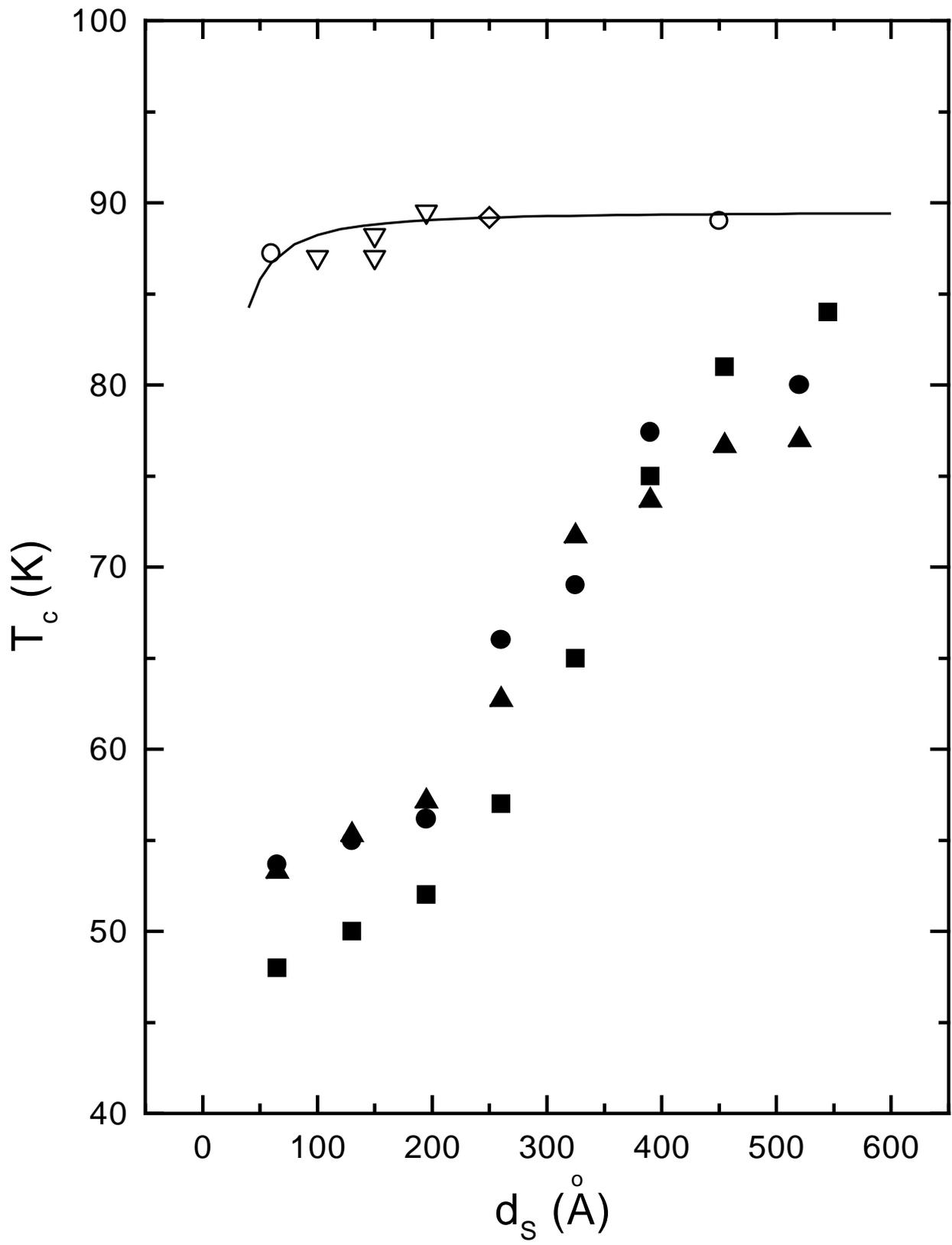
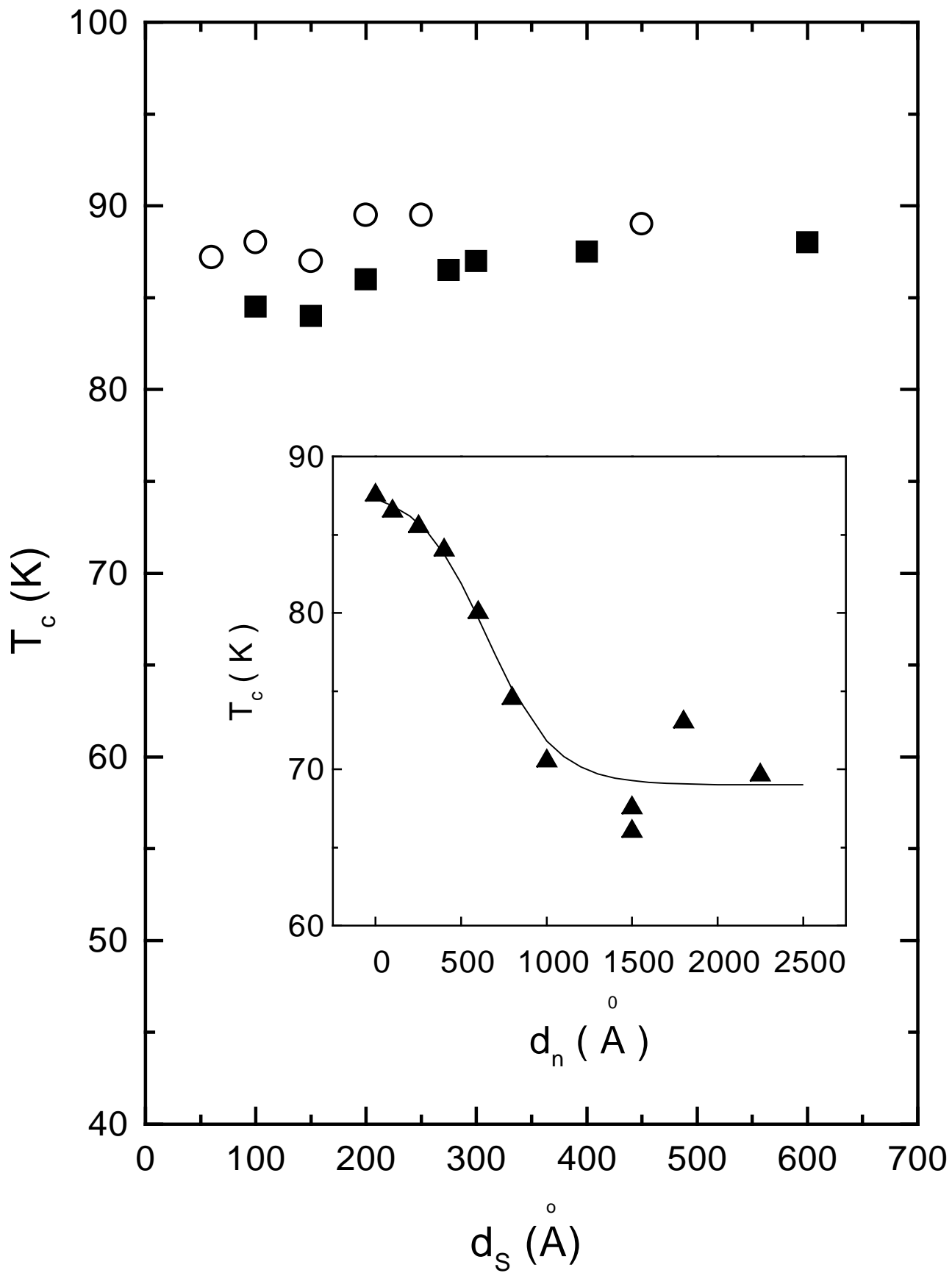
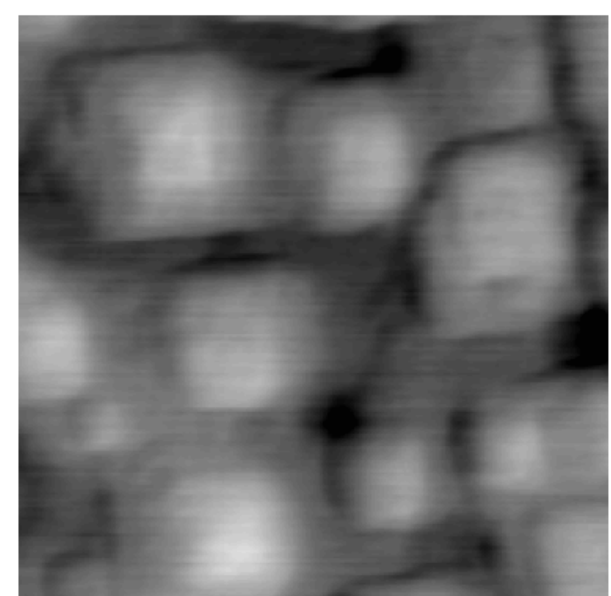
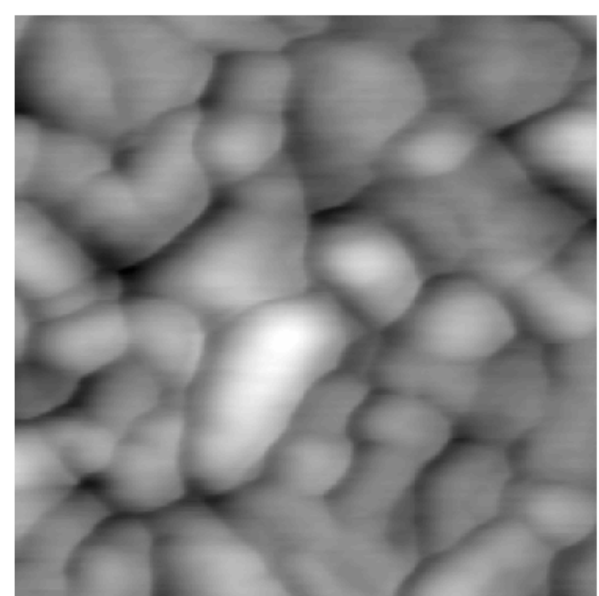
\end{document}